\documentclass[aps,prl,notitlepage,10pt,twocolumn,superscriptaddress]{revtex4-1}
\usepackage{dcolumn}
\usepackage{bm}
\usepackage{amsmath,amsthm,amssymb}
\usepackage{epstopdf}
\usepackage{graphicx}
\usepackage{color}
\usepackage{verbatim}

\newcommand{\ket}[1]{ |#1  \rangle}

\begin{document}
\title{Sub-Poissonian statistics of Rydberg-interacting dark-state polaritons}
\newcommand{\affA}{Physikalisches Institut, Universit\"at Heidelberg, Im Neuenheimer Feld 226, 69120 Heidelberg, Germany.}
\newcommand{\affB}{Max-Planck-Institut f\"ur Kernphysik, Saupfercheckweg 1, 69117 Heidelberg, Germany.}
\newcommand{\affC}{Institut f\"ur Theoretische Physik, Universit\"at Heidelberg, Philosophenweg 16, 69120 Heidelberg, Germany
}

\author{C. S. Hofmann}\thanks{These authors contributed equally to this work}\affiliation{\affA}
\author{G. G\"unter$^*$}\affiliation{\affA}
\author{H. Schempp$^*$}\affiliation{\affA}
\author{M. Robert-de-Saint-Vincent}\affiliation{\affA}
\author{M. G\"arttner}\affiliation{\affB}\affiliation{\affC}
\author{J. Evers}\affiliation{\affB}
\author{S. Whitlock}\email{whitlock@physi.uni-heidelberg.de}\affiliation{\affA}
\author{M. Weidem\"uller}\email{weidemueller@uni-heidelberg.de}
\affiliation{\affA}
\date{\today}

\begin{abstract}
 We observe individual dark-state polaritons as they propagate through an ultracold atomic gas involving Rydberg states coupled via an electromagnetically induced transparency resonance. Strong long-range interactions between Rydberg excitations give rise to a blockade between polaritons, resulting in large optical nonlinearities and modified polariton number statistics. By combining optical imaging and high-fidelity detection of the Rydberg polaritons we investigate both aspects of this coupled atom-light system. We map out the full nonlinear optical response as a function of atomic density and follow the temporal evolution of polaritons through the atomic cloud. In the blockade regime the statistical fluctuations of the polariton number drop well below the quantum noise limit. The low level of fluctuations indicates that photon correlations modified by the strong interactions have a significant back-action on the Rydberg atom statistics.
\end{abstract}

\maketitle
Interfacing light and matter at the quantum level is at the heart of modern atomic and optical physics and enables new quantum technologies involving the manipulation of single photons and atoms. A prototypical atom-light interface is electromagnetically induced transparency (EIT)~\cite{Fleischhauer2005}, which gives rise to hybrid states of photons and atoms called dark-state polaritons (DSPs)~\cite{Fleischhauer2000}. These long-lived quasi-particles simultaneously possess the properties of both the photonic and the atomic degrees of freedom, which can be interchanged in a fully coherent and reversible process. This has been intensively studied over the last decade, for instance, in the context of slow light~\cite{Hau1999,*Liu2001,*Phillips2001,*Bajcsy2003}, or to imprint a magnetic moment onto light fields~\cite{Karpa2006} and to realize giant electro-optical effects~\cite{Mohapatra2008}. 
Qualitatively new effects occur in \emph{strongly-interacting} atomic systems in which the atomic admixture can mediate polariton-polariton interactions, leading to highly nonlinear~\cite{Pritchard2010,Peyronel2012,Parigi2012} and nonlocal optical effects~\cite{Sevincli2011} as well as the emergence of correlations in both the atomic and the light fields~\cite{Petrosyan2011,Gorshkov2011}. The ability to produce and coherently control the propagation of quantum fields using interacting dark-state polaritons is expected to open up new applications including few photon nonlinear optics~\cite{Pritchard2010,Shahmoon2011,Gorshkov2011}, non-classical light sources~\cite{Petrosyan2011,Honer2011,Pritchard2013,Dudin2012,Maxwell2013,Peyronel2012}, photonic quantum logic gates~\cite{Friedler2005,Petrosyan2008,Maxwell2013}, and new types of strongly-interacting quantum gases~\cite{Chang2008,*Fleischhauer2008,*Zimmer2012}.

\begin{figure}[!t]
\includegraphics[width=0.95\columnwidth]{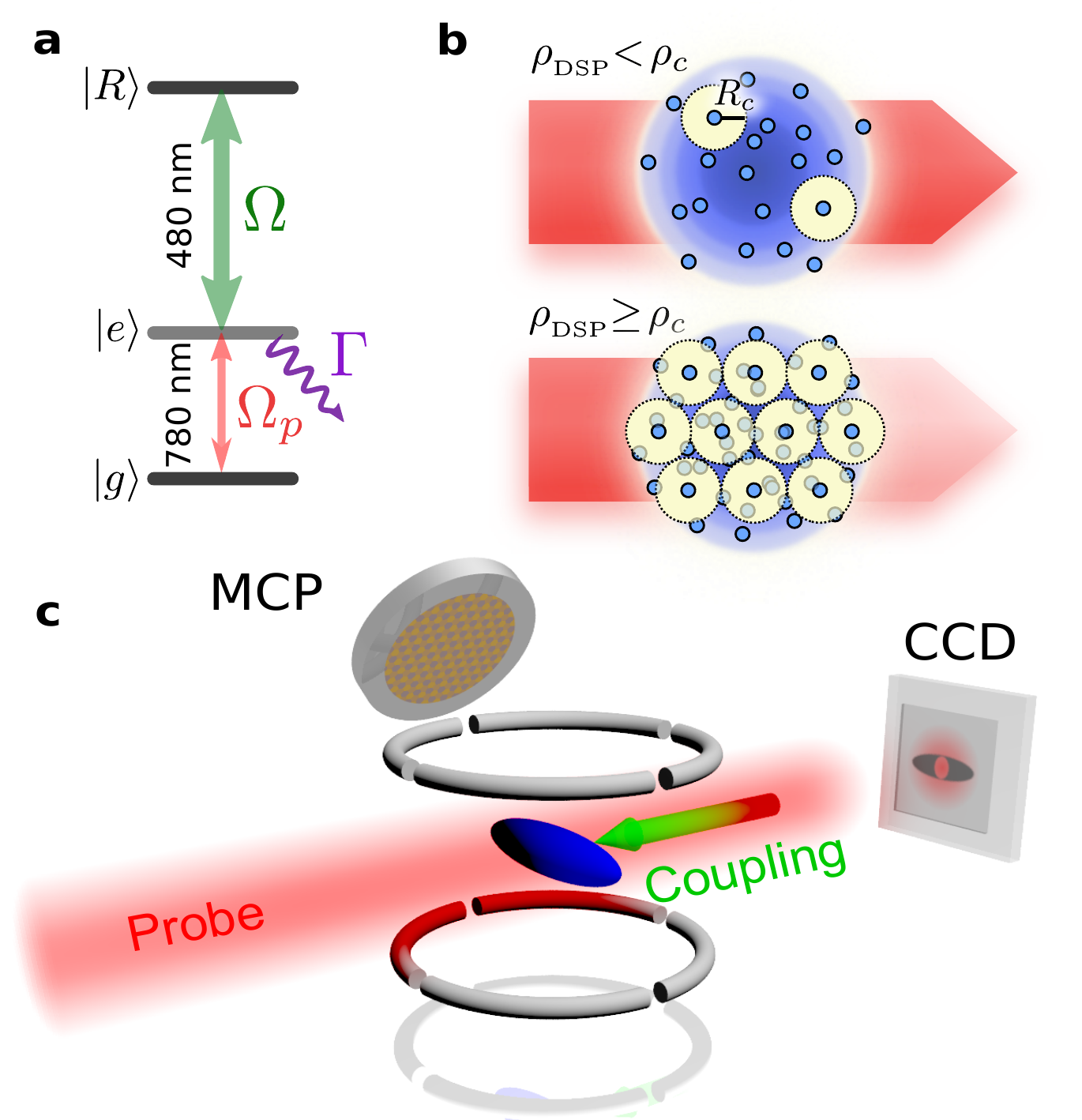}
\caption{\textbf{Experimental setup}. \textbf{a,} Level scheme used to observe electromagnetically induced transparency in a Rydberg interacting gas. \textbf{b,} Cross-section of the atomic cloud (blue) showing the effect of Rydberg blockade on polariton statistics and nonlinear absorption for small and large polariton densities. \textbf{c,} Schematic of the experimental setup. The atomic cloud is illuminated by the probe beam, yielding an absorption image on a CCD camera with a local transparency due to the coupling beam. Electrodes and a microchannel plate detector (MCP) allow for field ionisation and detection of Rydberg atoms. }\label{fig:schematic}
\end{figure}

In this Letter we investigate both light and matter aspects of strongly interacting DSPs in a dense atomic gas involving highly excited Rydberg states. We introduce a new way to study EIT which directly probes the matter-part of the polariton wavefunction and benefits from a high detection sensitivity to Rydberg atoms to provide complementary information to that accessible from the transmitted light. An important feature of this system is the Rydberg blockade~\cite{Lukin2001,Comparat2010} which gives rise to dissipative interactions and ultimately to correlations between DSPs~\cite{Gorshkov2011}. We observe the temporal dynamics of polaritons as well as the strongly nonlinear optical response and sub-Poissonian polariton statistics when entering the strongly interacting regime. Using a state-of-the-art theoretical approach for the coupled atom-light system we find good agreement with all aspects of the experiment except for the low level of sub-Poissonian fluctuations. This indicates that the atomic correlations are significantly stronger than can be accounted for by assuming the propagation of classical light fields, showing the recently observed modification of photon statistics due to Rydberg interactions~\cite{Peyronel2012} can also cause a significant back-action on the Rydberg atom statistics.

The basic principle of EIT involves a three-level system (Fig.~\ref{fig:schematic}a): two long-lived states $|g\rangle$ and $|R\rangle$ are coupled via a short-lived state $|e\rangle$ which spontaneously decays with rate $\Gamma$. Strong laser coupling $\Omega$ of the $|e\rangle \rightarrow\nobreak |R\rangle$ transition creates an Autler-Townes doublet of dressed states, such that the transition amplitudes for the $|g\rangle \rightarrow\nobreak |e\rangle$ resonance interfere destructively, giving rise to a narrow transparency resonance. In the weak probe limit, a single probe photon will be converted into a a dark-state polariton~\cite{Fleischhauer2000}, $\ket{D_1} = \cos(\theta)\ket{g^{(0)}} - \sin(\theta)\ket{R^{(1)}}$, where the collective quantum states involving $N$ atoms and the light field are $\ket{g^{(0)}} = \ket{g_1 ... g_N}\ket{1}$ and $\ket{R^{(1)}} = 1/\sqrt{N}\sum_{i=1}^{N} \ket{g_1 .. R_i .. g_N}\ket{0}$, where $|0\rangle,|1\rangle$ indicate $0$ or $1$ photons in the probe field. The mixing ratio $\tan(\theta)=\sqrt{{\rho \sigma c \Gamma}}/{\Omega}$ determines the properties of the polariton depending on the atom density $\rho$, the resonant scattering cross-section $\sigma=3\lambda^2/2\pi$, and the speed of light $c$. The DSP propagates through the atomic medium with negligible loss and reduced group velocity $v_g\approx c/\tan^2(\theta)$, before being coherently converted back to the original optical mode (Fig.\ref{fig:schematic}b). For increasing atomic density the probe field is compressed, thereby increasing the density of DSPs $\rho_{_\mathrm{DSP}}$. Since a DSP includes a large Rydberg state component, it strongly shifts the Rydberg states of nearby atoms within a distance $R_c$, breaking the EIT condition for additional excitations and leading to strong scattering of excess photons from the cloud~\cite{Petrosyan2011}. This imposes a critical density of DSPs $\rho_c\approx (4\pi R_c^3/3)^{-1}$ beyond which $\rho_{_\mathrm{DSP}}$ saturates (Fig.\ref{fig:schematic}b lower panel).

\begin{figure}[t!]
\includegraphics[width=0.95\columnwidth]{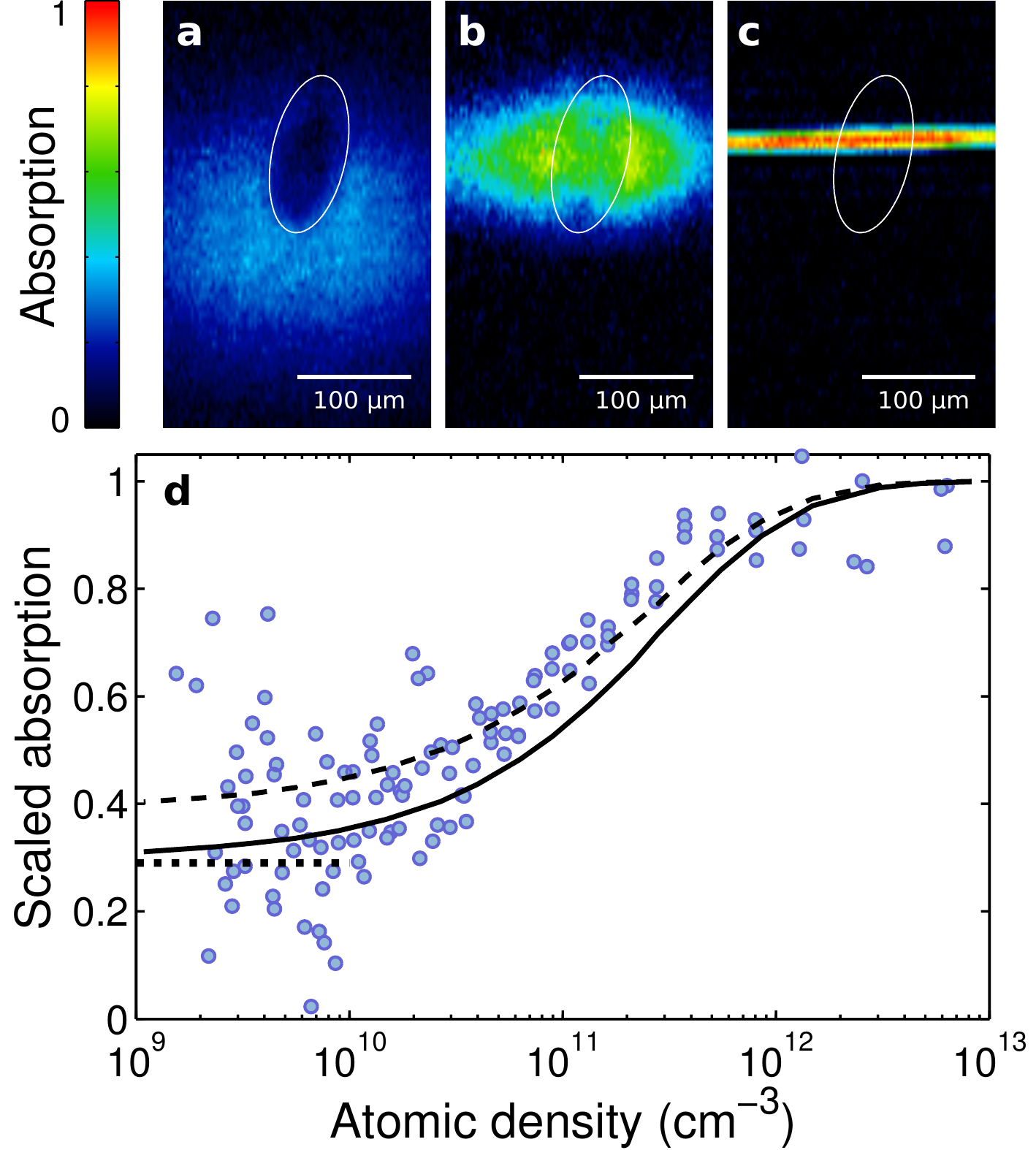}
\caption{\textbf{Nonlinear optical response of the Rydberg EIT medium}. Absorption images of the atomic cloud for three expansion times with peak densities corresponding to \textbf{a,} $5\times10^9$~cm$^{-3}$, \textbf{b,} $5\times10^{10}$~cm$^{-3}$ and \textbf{c,} $3\times10^{12}$~cm$^{-3}$. The probe pulse duration is $100~\mu$s. The EIT region illuminated by the coupling laser is indicated by white ellipses. \textbf{d} Absorption scaled to the two level response as a function of peak atomic density. The dotted horizontal line shows the expected low density EIT absorption due to the finite laser linewidths. The solid and dashed curves are the results of the Monte-Carlo simulations with $\gamma_{gR}$=1.7~MHz (solid) and $\gamma_{gR}$=2.6~MHz (dashed).}\label{fig:images}
\end{figure}

Our experimental setup is shown schematically in Fig.~\ref{fig:schematic}c. We use $^{87}$Rb atoms with states $|g\rangle=|5S_{1/2},F=2\rangle$, $|e\rangle=|5P_{3/2},F=3\rangle$ with $\Gamma=6.1$~MHz, and the Rydberg state $|R\rangle=|55S_{1/2}\rangle$. We prepare a cigar shaped cloud of approximately $5\times10^4$ atoms at a temperature of $5~\mu$K in an optical trap. The atomic cloud is uniformly illuminated from the side by a circularly-polarized, resonant probe pulse with intensity $I_p=5~\mu$W/cm$^2$. The transmitted light is imaged onto a CCD camera with an optical resolution of $\approx 12~\mu$m (Rayleigh criterion). The resonant coupling laser is counter aligned to the probe beam and passed through a diffractive optical element to create an approximately uniform intensity profile with $\Omega=5.1$~MHz over an elliptical region of $\approx 65\times 130~\mu$m$^2$ with an uncertainty of $10~\mu$m in each direction. The measured linewidths (including both decay and dephasing terms) for the probe transition and the two-photon transition (deduced from the EIT resonance) are $\gamma_{eg}\approx 6.4$~MHz and $\gamma_{gR}\approx 1.7~$MHz respectively, which gives an EIT coupling parameter $\mathcal{C}=\Omega^2/(\gamma_{eg}\gamma_{gR})=2.4$ and an EIT resonance width of $w=(1+\mathcal{C})\gamma_{gR}=5.8$~MHz. The anticipated blockade radius is $R_c = (2 C_6/w)^{1/6} \approx 5~\mu$m with the van der Waals interaction strength $C_6\approx 50$~GHz~$\mu$m$^6$. To control the atomic density before pulsing on the probe light we switch off the optical trap and vary the expansion time between $20~\mu$s and $4.5$~ms, varying the peak density (along the line-of-sight) between $\rho\approx 10^{13}-10^9$~cm$^{-3}$. This is coupled to a change in the cloud length in the probe direction between $\approx 2~\mu$m and $100~\mu$m and a change in the optical density of the cloud between $3$ (resolution limited) and $0.1$.

Figures~\ref{fig:images}(a-c) show absorption images of the atomic cloud for three densities. By analysing the images we measure the absorption in both the EIT region and the region where the coupling to the third level can be neglected (with susceptibility $\chi_{2L}$ where $2L$ refers to the two-level response). For low densities $\rho\lesssim 10^{10}$~cm$^{-3}$ (corresponding to long expansion times) we observe a high degree of transparency, which can be attributed to the low density of DSPs.  The measured scaled optical susceptibility $\chi_0/\chi_{2L}=0.3 \pm 0.1$ agrees with the expected on-resonant optical susceptibility in the non-interacting EIT regime $1/(1+\mathcal{C})\approx 0.29$. At densities above $\rho\approx 1\times10^{10}$~cm$^{-3}$ the transparency decreases due to interactions and above $\rho\approx 4\times10^{11}$~cm$^{-3}$ the EIT spot vanishes almost completely.

The full nonlinear optical response as a function of peak atomic density is shown in Fig.~\ref{fig:images}d. To model the data we use an optimised version~\cite{heeg2012} of the rate equation model originally introduced by Ates {\it et al.}~\cite{ates2007a,*Ates2011}. We assume random atom positions distributed according to the measured cloud shape and use Monte-Carlo sampling to solve for the steady state populations in states $|g\rangle,|e\rangle$ and $|R\rangle$ for each atom, taking into account the level shifts due to interactions with nearby Rydberg atoms. We find that the effect of nonlinear absorption as the probe propagates through the cloud must be accounted for. Therefore, to solve for the steady state populations for any given atom, we recursively solve for the local probe intensity which is attenuated by all preceding atoms, which in turn depends on their steady state populations (see Supplemental Material).

We find good agreement between the measured optical response and the results of the Monte-Carlo simulations using the calculated interaction strength and all other input parameters determined from independent measurements (Fig.~\ref{fig:images}d solid curve). The simulations closely reproduce the observed scaling of the nonlinear absorption, however at intermediate and high densities the scaled absorption is underestimated slightly. The discrepancy could be explained by the combination of experimental uncertainties in the measured dephasing rates or additional density-dependent dephasing. In the low-density limit the simulations with the measured value of $\gamma_{gR}=1.7$~MHz (solid line) are consistent with the experimental data. For comparison we show a fit to the data assuming a larger overall dephasing of $\gamma_{gR}$=2.6~MHz (dashed curve) which would still be consistent with the experimental uncertainties. 

\begin{figure}
\includegraphics[width=0.95\columnwidth]{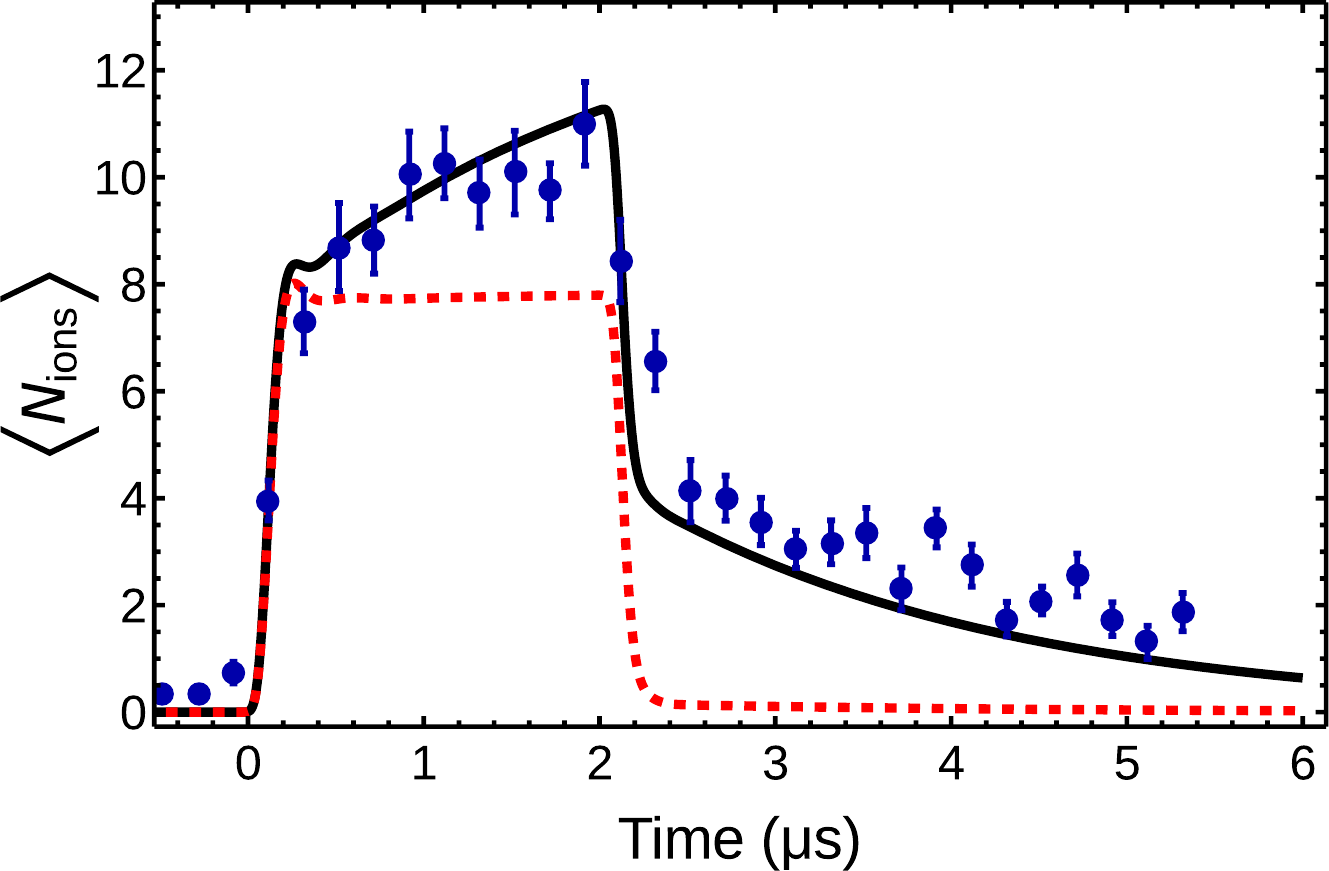}
\caption{\textbf{Polariton dynamics observed during a $2~\mu$s probe pulse measured via the field ionized Rydberg state population.} Each data point is computed from the average of 20 runs of the experiment. The vertical error bars represent the statistical error of the mean.  The dashed red line shows the Rydberg population for the three level system as computed from the time-dependent solution of the single-atom optical Bloch equations. The solid line is a fit to the data including a coupling to both Rydberg Zeeman sublevels $m_j=\pm 1/2$ which accounts for the slow evolution and residual Rydberg population after $2~\mu$s.}\label{fig:polariton}
\end{figure}

In our experiments the DSPs are almost entirely matter-like ($\cos^2(\theta)\approx 10^{-3}-10^{-7}$). Therefore, by measuring the Rydberg population we make a projective measurement of the number of polaritons inside the cloud. This is achieved by field ionizing the Rydberg states and subsequently detecting the individual ions on a microchannel plate (MCP) detector (see Supplemental Material). Figure~\ref{fig:polariton} shows the measured time evolution of the number of detected ions during a $2.2~\mu$s probe pulse in the non-interacting regime. At short times we observe a rapid increase of the detected ions reflecting the entry of DSPs into the atomic cloud, which then approaches a steady state value of around $\langle N_{\mathrm{ions}}\rangle\approx 10$. Accounting for our ion detection efficiency ($\eta\approx 0.4$) this corresponds to $\approx 25$ polaritons inside the cloud. After the pulse the Rydberg state population drops rapidly, which is expected as the polaritons are coherently mapped back onto the light field. The dashed line in Fig.~\ref{fig:polariton} shows the expected evolution found by solving the three-level optical Bloch equations. The additional slow rise during the pulse and the residual population after the pulse observed in the experiment can be attributed to laser coupling to additional Zeeman sublevels as seen in the solution to the four level optical Bloch equations (Fig.~\ref{fig:polariton} solid line).

\begin{figure}
\includegraphics[width=0.95\columnwidth]{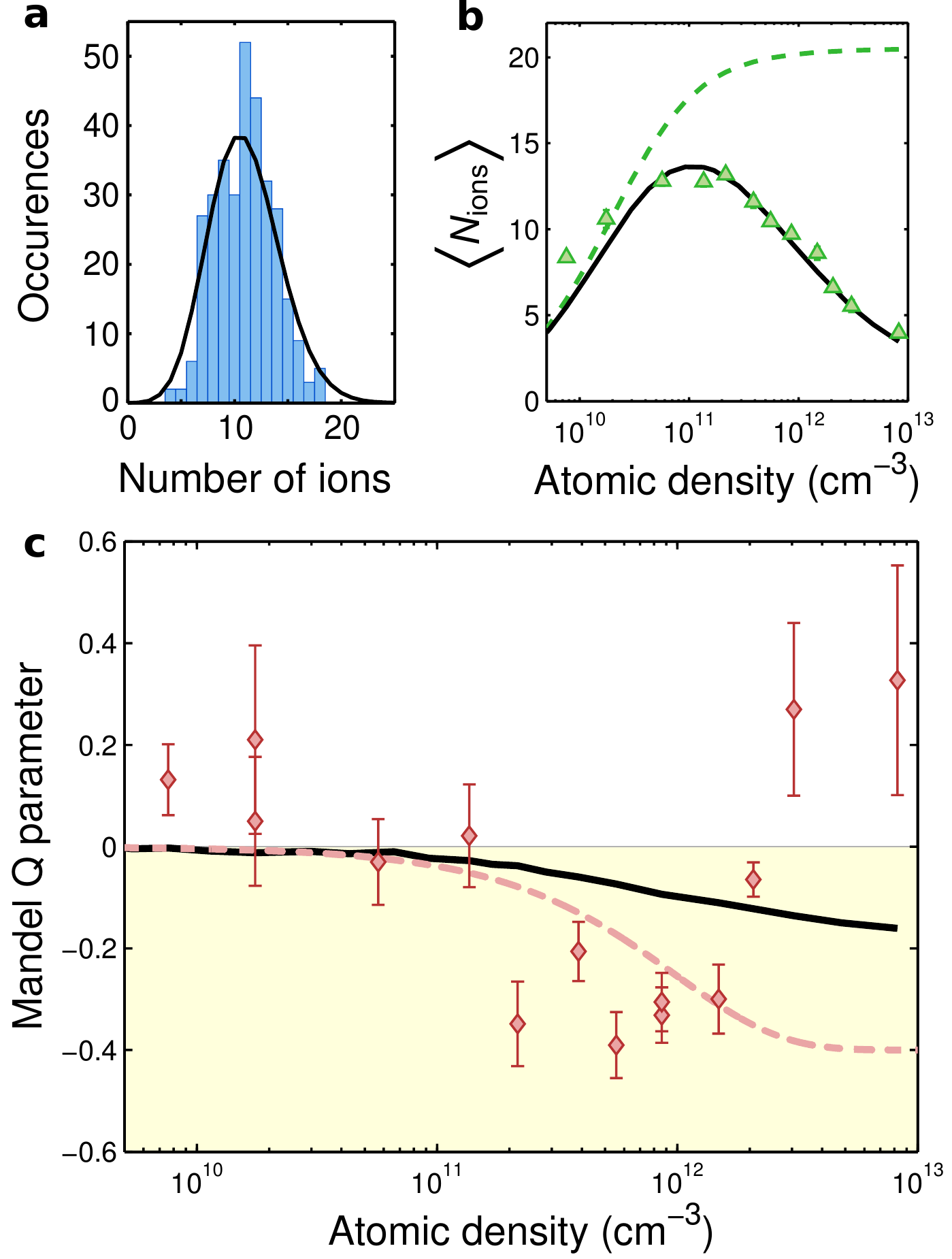}
\caption{\textbf{Statistical distribution of the detected ions during a $2~\mu$s probe pulse showing the transition to sub-Poissonian statistics.} \textbf{a,} Ion number histogram for $\rho=5\times 10^{11}$~cm$^{-3}$ with $Q=-0.3$ compared to a Poisson distribution with the same mean (solid line). \textbf{b,} Mean number of ions as a function of atomic density which decreases above $\rho=10^{11}$~cm$^{-3}$ due to blockade. The solid line is the result of the Monte-Carlo calculations with $\eta=0.4$. The dashed line shows the expected number neglecting interactions and probe absorption, which is proportional to the total number of atoms in the excitation volume. \textbf{c,} Mandel $Q$ parameter as a function of atomic density. Statistical errors are estimated by dividing the data for each density into $3$ subsets of 50 experimental runs each and then calculating the standard error of the mean $Q$ parameter. The solid line is the result of Monte-Carlo calculations using the same parameters as in Fig.~\ref{fig:images}d and Fig.~\ref{fig:qfactor}b. The dashed curve is the hard sphere model (see text) assuming an initial Poissonian photon number distribution with $\rho_{crit}=1\times 10^{12}$~cm$^{-3}$. }\label{fig:qfactor}
\end{figure}

To investigate the appearance of correlations in the strongly interacting regime, we map out the distribution of DSPs inside the cloud as a function of atomic density. After $2~\mu$s of probe laser illumination we field ionize the Rydberg atoms and count the number of detected ions. This is repeated between 150-300 times for each density to construct histograms of the ion number distribution (Fig.~\ref{fig:qfactor}a) and to extract the mean and variance. In Fig.~\ref{fig:qfactor}b we show the measured mean ion number $\langle N_{\mathrm{ions}}\rangle$. The low-density behaviour is determined by the changing number of atoms in the excitation volume as a function of expansion time (dashed line). For $\rho>10^{11}$~cm$^{-3}$ we observe a rapid decrease of $\langle N_{\mathrm{ions}}\rangle$ due to probe laser attenuation and polariton-polariton interactions. The solid line shows the results of the Monte-Carlo simulation using the same parameters as for Fig.~\ref{fig:images}. To obtain the best agreement we adjust only two parameters: the detection efficiency $\eta=0.4$ and the coupling laser ellipse axes of $65~\mu$m and $130~\mu$m respectively in agreement with the experimental values.

Simultaneous to the onset of the polariton blockade we also observe a significant narrowing of the statistical distribution of the detected ions (Fig.~\ref{fig:qfactor}a). We quantify this using the Mandel $Q$ parameter $Q=\nobreak \mathrm{var}(N_{ions})/\langle N_{ions}\rangle-\nobreak 1$ which compares the variance of an observed distribution to a Poisson distribution with the same mean. For a Poissonian process $Q = 0$, whereas a sub-Poissonian process yields $Q < 0$ reflecting the emergence of spatial and temporal correlations~\cite{Ates2006,Reinhard2008,Viteau2012}. Figure 4c shows the measured Mandel $Q$ parameter as a function of the peak density of the atomic cloud. In the low-density limit $Q\approx 0$ which is expected for a coherent input light field. In the density range $2\times 10^{11}$~cm$^{-3}<\rho<2\times 10^{12}$~cm$^{-3}$ we observe a clear transition to sub-Poissonian statistics with $Q=-0.32\pm0.04$. Accounting for imperfect ion detection indicates that the true distribution of DSPs is much narrower, corresponding to strong spatial and temporal correlations (since $Q_{\mathrm{DSP}}=Q/\eta$) ~\cite{Ates2006,Reinhard2008,Viteau2012}. 

Despite the good agreement of the Monte-Carlo model with both the optical response data (Fig.~\ref{fig:images}d) and the mean number of polaritons (Fig.~\ref{fig:qfactor}b), it clearly fails to describe the experimentally measured $Q$ parameters which are much lower than predicted (Fig.~\ref{fig:qfactor}c). We attribute this to the neglected effect of photon correlations (the Monte-Carlo algorithm assumes a classical field at each step characterised by a Rabi frequency $\Omega_p$) which can build up as a consequence of dissipative polariton-polariton interactions~\cite{Peyronel2012}. A complete many-body treatment of this problem, including the effects of the quantized electromagnetic field is still lacking. However, a simple estimate for the possible effects of photon correlations can be made assuming hard-sphere interactions. We discretize the excitation volume into elements corresponding to the size of a blockade sphere and assume an initial Poisson distribution for the number of photons per element with a mean of $\langle N_{ph}\rangle \approx \Omega_p^2\rho/(\Omega^2\rho_c)$~\cite{Petrosyan2011}. For $N_{ph}\geq 1$ we assume that one DSP is created with unity probability and the excess probe photons couple to bright states which include the intermediate state. For our experimental parameters the transition point corresponds to $\rho_{crit}\approx 1\times 10^{12}$ cm$^{-3}$. This simple model gives $Q\approx\eta \exp(-\rho/\rho_{crit})-\eta$, which is in closer agreement with the data, confirming that the development of photon number correlations within the cloud plays an important role on the DSP statistics (Fig.~\ref{fig:qfactor}c). 

Above $\rho\approx 3\times 10^{12}$ we find that the measured $Q$ values increase to super-Poissonian values. The cause has not yet been identified, but we note that it coincides with the point when the cloud becomes quasi-1D with respect to the Rydberg excitations, which might modify the effects of interactions on the propagating light field. Alternatively, it could be due to the production of a small number of spontaneously created ions (below our detection sensitivity)~\cite{Vincent2012} which could dramatically affect the polariton statistics.

Our experiments provide a direct observation of strongly-interacting DSPs realized using EIT in an ultracold Rydberg gas. It is instructive to compare our matter-based observations of DSPs with the observation of photonic correlations in the outgoing field. Two-photon correlations have recently been observed in the second order intensity correlation function of light $g^{(2)}(\tau)$ retrieved from small Rydberg ensembles~\cite{Dudin2012,Maxwell2013,Peyronel2012}. In comparison, our method would allow for studying the full polariton counting statistics, thereby also giving access to higher order correlations~\cite{Ates2006}. Furthermore, we point out that in general there is no direct mapping between the correlations on the light field and the atomic correlations. For instance, in a dense but optically thin medium, excess photons may not be efficiently scattered out of the forward direction thereby washing out the $g^{(2)}$ contrast, despite strong atomic correlations. Alternatively, spin-wave dephasing due to atomic interactions can alter the statistics of the emitted light without blockade~\cite{Dudin2012,Dudin2012b,Bariani2012}, even though the atomic correlations could be nearly classical. Studying both aspects enables stringent tests of theoretical models and provides a complete picture of EIT in strongly-interacting systems. Ultimately, this will open an avenue to create new quantum technologies such as on-demand sources of single photons and other nonclassical states of light, or to engineer new types of strongly-correlated systems.

\begin{acknowledgements}\vskip 5pt
\noindent\footnotesize {\bf{Acknowledgements}} We acknowledge N. L. M. M\"uller, A. Faber and H. Busche for contributions to the experimental apparatus, K. P. Heeg for work on the theoretical model and T. Pohl for valuable discussions. This work is supported in part by the Heidelberg Center for Quantum Dynamics and supported by the Deutsche Forschungsgemeinschaft under WE2661/10.2. M.R.D.S.V. acknowledges support from the EU Marie-Curie program (Grant No. FP7-PEOPLE-2011-IEF-300870).

\end{acknowledgements}

\bibliography{bib_universal,polariton_statistics}

\end{document}